# ON THE DEVELOPMENT OF METHODOLOGY FOR PLANNING AND COST-MODELING OF A WIDE AREA NETWORK

B. Ahmedi[1] and P. Mitrevski[2]

[1]State University of Tetovo, Faculty of Mathematics and Natural Sciences,
Tetovo, Macedonia

[2]University of St. Clement Ohridski, Faculty of Technical Sciences,
Bitola, Macedonia

*ABSTRACT*

*The most important stages in designing a computer network in a wider geographical area include: definition of requirements, topological description, identification and calculation of relevant parameters (i.e. traffic matrix), determining the shortest path between nodes, quantification of the effect of various levels of technical and technological development of urban areas involved, the cost of technology, and the cost of services. These parameters differ for WAN networks in different regions – their calculation depends directly on the data "in the field": number of inhabitants, distance between populated areas, network traffic density, as well as available bandwidth. The main reason for identification and evaluation of these parameters is to develop a model that could meet the constraints imposed by potential beneficiaries. In this paper, we develop a methodology for planning and cost-modeling of a wide area network and validate it in a case study, under the supposition that behavioral interactions of individuals and groups play a significant role and have to be taken into consideration by employing either simple or composite indicators of socioeconomic status.*

## 1. INTRODUCTION

A Wide Area Network (WAN) is a computer network that covers a group of LAN networks that are administered separately and have different technology [1,2,3]. A network in a wider geographic area actually represents the infrastructure through which most ordinary services are carried out, all the way up to the most serious operations, such as monetary transactions. Creating an effective network that meets the needs of a particular region with minimal cost is still a challenge [4,5]. The end-user experience at the branch can be enhanced and the number of transactions across a WAN can be minimized by the use of WAN optimization/acceleration devices [6]. In 2009, computer academics at the University of California, Berkeley, asked whether it was more cost-effective to compute *locally* or reach across the network to powerful *remote* or distributed systems. It turned out the costs for storage and servers have fallen faster than that of network connections. Although WAN costs fell by a factor of 2.7x, computational equipment fell by a stunning 16 times. Ultimately, IT has to decide if the cloud makes economic sense. All the factors have to be considered when moving to cloud [7]: a company can write off all its assets and go to a fixed monthly cost – but, if the cloud environment is designed correctly, it should not see that much cost for bandwidth [8]. Nevertheless, WANs and the Internet were getting clogged long before the cloud: multimedia, Web and video conferencing, surfing, VoIP and unified communications already stress those networks that haven't been thoroughly boosted.

A WAN designer is required to possess a deep knowledge of technologies that will be encompassed in the process of building a WAN, their performance, price and connectivity. One also needs to consider the burden of network traffic and the distance of data transfer in certain areas and at certain times. A fundamental question that arises is *how to calculate (quantitatively) the traffic in these regional networks where the communication volume is variable*. Therefore, to construct an efficient WAN network, preceding calculations are required, such as identification and calculation of the traffic, the distances between cities, the number of participants in the traffic, the time of use of network infrastructure of participants, etc. Finally, it is also about the calculation of *carried traffic* (in Erlangs), as well as the *bandwidth* of the edges of the network graph.

In certain geographic areas, and at certain times, the traffic volume in a computer network is variable [9]. The *traffic matrix* describes the traffic map from one network location to all other locations, and it can be structured so that it represents the average value of the amount of traffic carried between locations – *nodes* of the network. Moreover, the introduction of either *simple* or *composite* indicators of *socioeconomic status* (i.e. *a socioeconomic indicator* – SEI), as a concrete contribution in this paper, helps obtaining more accurate information, needed in the development of methodology for planning and cost-modeling of a WAN network.

The remainder of this paper is organized as follows. Section 2 focuses on related work and identifies the main motivations. Section 3 discusses most of the necessary steps for effective construction of a WAN – calculation of traffic matrix, algorithms for finding the shortest path, as well as calculating traffic and bandwidth between nodes in a case-study network: we assume that 21 cities in the Republic of Macedonia are the graph nodes with 30 internode edges, and the weights of these edges are a measure for the distance between cities; in addition, to determine the shortest paths, *Dijkstra's* and *Floyd-Warshall* algorithms are applied. Prices of different technologies are evaluated in Section 4, where we shift our attention to cost-modeling. Finally, Section5 concludes the paper.

## 2. MOTIVATION AND RELATED WORK

Al-Wakeel's research report [9] is aimed towards cost study and analysis of WANs and Internetworks design. It focuses on the *economic* and *performance* characteristics of various network technologies and carrier service options, and evaluates the conditions in which each of these technologies is optimal. A top-down, step-by-step process has been developed and quantitative, business-oriented cost models for the network design have been built, in order to develop planning programs for constructing a WAN that is cost-effective for a large organization.

De Montis *et al.* [10] study the structure of a network representing the interurban *commuting traffic* of the Sardinia region, Italy, and use a *weighted network representation* where vertices correspond to towns and the edges to the actual commuting flows. However, they discuss the interplay between the topological and dynamical properties of the network, as well as their relation with *socio-demographic variables* such as *population* and *monthly income*. This analysis provides analytical tools for a wide spectrum of applications, and some of the principles laid out could easily be mapped to the problem of planning and cost-modeling of a wide area *computer network*, as well.

International Journal of Computer Networks & Communications (IJCNC) Vol.6, No.3, May 2014

Taylor *et al.* [11] aim to identify associations between *demographic* and *socioeconomic factors* and *home Internet use patterns* in the Central Queensland region, Australia. This research hypothesizes that there are differences in Internet usage patterns between: young and old, male and female, people in urban and rural areas, married and unmarried, well-educated and less educated, rich and poor, and employed and unemployed. This paper examines differences in home Internet use across these parameters and the associations between home Internet consumption patterns and *demographic* and *socioeconomic factors*.

These are the main motivations for the introduction of *socioeconomic measures* (which are regularly used in official statistics to illustrate patterns of behavior and outcomes, and to support and develop policies) in the development of a model which will determine the exact cost of a wide area computer network that performs within the limits set by the demands of prospective users.

## 3. ON THE DEVELOPMENT OF METHODOLOGY FOR PLANNING OF A WIDE AREA NETWORK – A CASE STUDY

### 3.1. Calculation of the Traffic Matrix

Calculation of the traffic matrix is based on the number of households in the cities, time of use of the networks and the number of service users over a network. The term "household" is considered a family or other community of people who declare that they are living together and spend their income. In other words, under the term household we consider so-called "collective household", composed of persons living permanently in institutions for care of children and adults. In a city, there are two types of households: residential and commercial. The number of households in each node of the graph is calculated according to the following formula [9]:

$$T = \frac{P}{N} \quad (1)$$

where:
– number of households in the city
– number of inhabitants in the city
N – number of inhabitants in a house (e.g.: this number in the Republic of Macedonia is 3.5) [9,12]

The total traffic in Erlangs (E) for any city (Table 2) is calculated as:

$$TF = \left( C \cdot \frac{P}{N} \cdot \frac{CC \cdot CL}{24} + R \cdot \frac{P}{N} \cdot \frac{CR \cdot RL}{24} \right) \cdot SEI \quad (2)$$

where:
C – commercial households (in percent)
CC – number of active on-line sessions per commercial household per day (24 hours)
CL – commercial session duration (in hours)
R – residential households (in percent)
CR – number of active on-line sessions per residential household per day (24 hours)
RL – residential session duration (in hours)
SEI – composite socioeconomic indicator.

The level of usage of a WAN network in a city depends not only on the population of that city, but also on their demographic structure and behavioral interactions of individuals and groups. It is quantified by a composite socioeconomic indicator (SEI), as a linear sum of products of weights $w_i$, and several measures of labor status and education, $a_i$ ($0 \leq a_i \leq 1$) (Table 1). The weights $w_i$ are defined on the basis of statistical reports [12]:

$$SEI = \sum_{i=1}^{k} w_i a_i \qquad (3)$$

where:
$a_1$ – percentage of public employees in that city;
$a_2$ – percentage of employees in enterprises in that city;
$a_3$ – percentage of pupils in that city;
$a_4$ – percentage of students in that city;
$a_5$ – percent of unemployed in that city;
$a_6$ – percent of the other inhabitants; i.e.:

$$a_1 = \frac{I_1}{P},\ a_2 = \frac{I_2}{P},\ a_3 = \frac{I_3}{P},\ a_4 = \frac{I_4}{P},\ a_5 = \frac{I_5}{P},\ a_6 = \frac{I_6}{P} \qquad (4)$$

where:
$I_1$ – number of employees in the public sector in that city;
$I_2$ – number of employees in enterprises in that city;
$I_3$ – number of pupils in that city;
$I_4$ – number of students in that city;
$I_5$ – number of unemployed in that city;
$I_6$ – number of other population;
$P$ – total population of the city.

$w_1$ – use of ICT and Internet in the public sector;
$w_2$ – use of ICT and Internet in enterprises;
$w_3$ – use of ICT and Internet by pupils;
$w_4$ – use of ICT and Internet by students;
$w_5$ – use of ICT by unemployed;
$w_6$ – use of ICT by others.

The traffic *between* cities A and B is calculated according to the formula:

$$T_{AB} = T_A \cdot \frac{P_B}{P_T} \qquad (5)$$

where:
$T_A$ – the traffic between city A and all other cities;
$P_B$ – number of residents in city B;
$P_T$ – number of all the inhabitants of 21 cities.

The SEI indicator has a significant impact in the calculation of the traffic between cities. The biggest users of computer networks in Macedonia are students and pupils (96.4%) and from here it turns out that the calculation of the traffic matrix between cities with SEI indicator included, traffic noticeably changes. If one carefully examines Table 3, there is a traffic increase between the nodes in the graph representing cities that have higher percentages of students, such as: Skopje–Bitola, Skopje–Tetovo, Skopje–Shtip, whereas there are no significant changes in traffic in Debar–Krusevo and Gevgelija–Kocani links, to name a few.

Table 1. Calculating the SEI.

| City | Households | Population | w1 | w2 | w3 | w4 | w5 | w6 | N | employees | I1 no.public sector (10%) | $a_1 = \frac{I_1}{N}$ | I2 no.business sector (60%) | $a_2 = \frac{I_2}{N}$ | I3 elementary students | $a_3 = \frac{I_3}{N}$ High School students | | I4 no. students | $a_4 = \frac{I_4}{N}$ | I5 no.unemployed | $a_5 = \frac{I_5}{N}$ | I6 no.other | $a_6 = \frac{I_6}{N}$ | $SEI = \sum_{i=1}^{6} w_i a_i$ |
|---|---|---|---|---|---|---|---|---|---|---|---|---|---|---|---|---|---|---|---|---|---|---|---|---|
| | | | | | level of the republic | | | | | | | | | | | | | | | | | | | |
| 1 Struga | 14485 | 63376 | 0,720 | 0,360 | 0,964 | 0,964 | 0,25 | 0,25 | 63376 | 47811 | 19124,4 | 0,302 | 28686,6 | 0,453 | 6006 | 3229 | 0,146 | | 0 | 3034 | 0,048 | 3296 | 0,052 | 0,546 |
| 2 Ohrid | 16012 | 55749 | 0,720 | 0,360 | 0,964 | 0,964 | 0,25 | 0,25 | 55749 | 39074 | 15629,6 | 0,280 | 23444,4 | 0,421 | 4906 | 2605 | 0,135 | 1742 | 0,031 | 3809 | 0,068 | 3613 | 0,065 | 0,547 |
| 3 Bitola | 28942 | 95385 | 0,720 | 0,360 | 0,964 | 0,964 | 0,25 | 0,25 | 95385 | 66955 | 26782 | 0,281 | 40173 | 0,421 | 7421 | 4604 | 0,126 | 2973 | 0,031 | 6838 | 0,072 | 6594 | 0,069 | 0,541 |
| 4 Debar | 3917 | 19542 | 0,720 | 0,360 | 0,964 | 0,964 | 0,25 | 0,25 | 19542 | 13306 | 5322,4 | 0,272 | 7983,6 | 0,409 | 2306 | 713 | 0,154 | | 0 | 1211 | 0,062 | 2006 | 0,103 | 0,533 |
| 5 Kicevo | | 47700 | 0,720 | 0,360 | 0,964 | 0,964 | 0,25 | 0,25 | 47700 | 34354 | 13741,6 | 0,288 | 20612,4 | 0,432 | 3997 | 2327 | 0,133 | 599 | 0,013 | 3238 | 0,068 | 3185 | 0,067 | 0,537 |
| 6 Krushevo | 2706 | 9684 | 0,720 | 0,360 | 0,964 | 0,964 | 0,25 | 0,25 | 9684 | 7121 | 2848,4 | 0,294 | 4272,6 | 0,441 | 804 | 298 | 0,114 | | 0 | 837 | 0,086 | 624 | 0,064 | 0,518 |
| 7 Prilep | 24398 | 76768 | 0,720 | 0,360 | 0,964 | 0,964 | 0,25 | 0,25 | 76768 | 50641 | 20256,4 | 0,264 | 30384,6 | 0,396 | 6741 | 3977 | 0,140 | 1152 | 0,015 | 6760 | 0,088 | 7497 | 0,098 | 0,528 |
| 8 Gostivar | 18091 | 81042 | 0,720 | 0,360 | 0,964 | 0,964 | 0,25 | 0,25 | 81042 | 60136 | 24054,4 | 0,297 | 36081,6 | 0,445 | 7083 | 4844 | 0,147 | | 0 | 3546 | 0,044 | 5433 | 0,067 | 0,544 |
| 9 Tetovo | 20094 | 86580 | 0,720 | 0,360 | 0,964 | 0,964 | 0,25 | 0,25 | 86580 | 41528 | 16611,2 | 0,192 | 24916,8 | 0,288 | 10114 | 9795 | 0,230 | 6670 | 0,077 | 9761 | 0,113 | 8712 | 0,101 | 0,591 |
| 10 Skopje | 146566 | 506926 | 0,720 | 0,360 | 0,964 | 0,964 | 0,25 | 0,25 | 506926 | 3E+05 | 138894,4 | 0,274 | 208341,6 | 0,411 | 53586 | 27692 | 0,160 | 32911 | 0,065 | 15874 | 0,031 | 29627 | 0,058 | 0,585 |
| 11 Kumanovo | 27984 | 105484 | 0,720 | 0,360 | 0,964 | 0,964 | 0,25 | 0,25 | 105484 | 68606 | 27442,4 | 0,260 | 41163,6 | 0,390 | 11303 | 6514 | 0,169 | 1732 | 0,016 | 6472 | 0,061 | 10857 | 0,103 | 0,548 |
| 12 Kriva Palanka | 6600 | 20820 | 0,720 | 0,360 | 0,964 | 0,964 | 0,25 | 0,25 | 20820 | 13503 | 5401,2 | 0,259 | 8101,8 | 0,389 | 1718 | 843 | 0,123 | | 0 | 3154 | 0,151 | 1602 | 0,077 | 0,503 |
| 13 Sveti Nikole | 5698 | 18497 | 0,720 | 0,360 | 0,964 | 0,964 | 0,25 | 0,25 | 18497 | 12767 | 5106,8 | 0,276 | 7660,2 | 0,414 | 1454 | 661 | 0,114 | 205 | 0,011 | 1843 | 0,100 | 1567 | 0,085 | 0,515 |
| 14 Shtip | 15065 | 47796 | 0,720 | 0,360 | 0,964 | 0,964 | 0,25 | 0,25 | 47796 | 34171 | 13668,4 | 0,286 | 20502,6 | 0,429 | 4147 | 2794 | 0,145 | 1500 | 0,031 | 2434 | 0,051 | 2750 | 0,058 | 0,558 |
| 15 Veles | 16959 | 55108 | 0,720 | 0,360 | 0,964 | 0,964 | 0,25 | 0,25 | 55108 | 38479 | 15391,6 | 0,279 | 23087,4 | 0,419 | 4579 | 2763 | 0,133 | 375 | 0,007 | 2995 | 0,054 | 5917 | 0,107 | 0,527 |
| 16 Kocani | 11981 | 38092 | 0,720 | 0,360 | 0,964 | 0,964 | 0,25 | 0,25 | 38092 | 24041 | 9616,4 | 0,252 | 14424,6 | 0,379 | 3134 | 1719 | 0,127 | 2858 | 0,075 | 2884 | 0,076 | 3456 | 0,091 | 0,555 |
| 17 Radovish | 8270 | 28244 | 0,720 | 0,360 | 0,964 | 0,964 | 0,25 | 0,25 | 28244 | 19009 | 7603,6 | 0,269 | 11405,4 | 0,404 | 2661 | 894 | 0,126 | 2119 | 0,075 | 1593 | 0,056 | 1968 | 0,070 | 0,564 |
| 18 Negotino | 5898 | 19212 | 0,720 | 0,360 | 0,964 | 0,964 | 0,25 | 0,25 | 19212 | 14431 | 5772,4 | 0,300 | 8658,6 | 0,451 | 1817 | 775 | 0,135 | | 0 | 1085 | 0,056 | 1104 | 0,057 | 0,537 |
| 19 Kavadaci | 12026 | 38741 | 0,720 | 0,360 | 0,964 | 0,964 | 0,25 | 0,25 | 38741 | 27702 | 11080,8 | 0,286 | 16621,2 | 0,429 | 3438 | 1860 | 0,137 | | 0 | 3513 | 0,091 | 2228 | 0,058 | 0,529 |
| 20 Strumica | 15896 | 54676 | 0,720 | 0,360 | 0,964 | 0,964 | 0,25 | 0,25 | 54676 | 33452 | 13380,8 | 0,245 | 20071,2 | 0,367 | 5491 | 4018 | 0,174 | | 0 | 2344 | 0,043 | 9371 | 0,171 | 0,530 |
| 21 Gevgelija | 7221 | 22988 | 0,720 | 0,360 | 0,964 | 0,964 | 0,25 | 0,25 | 22988 | 15561 | 6224,4 | 0,271 | 9336,6 | 0,406 | 1869 | 991 | 0,124 | 2160 | 0,094 | 1459 | 0,063 | 948 | 0,041 | 0,578 |
| Σ | 345453 | 1492410 | | | | | | | | | | | | | | | | | | | | | | |

Table 2. The total traffic from 21 cities with SEI included.

|  | City | population | Households N= 3,5 T | CT(15%) | RT(85%) | TC (CT*CL* 4)/24, CL=1/2 h | TR (RT*RL* 1)/24, RL=1/2 h | TC+TR | SEI | (TC+TR)*SEI |
|---|---|---|---|---|---|---|---|---|---|---|
| 1 | Struga | 63376 | 18107,43 | 2716,11 | 15391,31 | 226,34 | 320,65 | 547,00 | 0,545661 | 298,474 |
| 2 | Ohrid | 55749 | 15928,29 | 2389,24 | 13539,04 | 199,10 | 282,06 | 481,17 | 0,546533 | 262,974 |
| 3 | Bitola | 95385 | 27252,86 | 4087,93 | 23164,93 | 340,66 | 482,60 | 823,26 | 0,540561 | 445,024 |
| 4 | Debar | 19542 | 5583,43 | 837,51 | 4745,91 | 69,79 | 98,87 | 168,67 | 0,533251 | 89,941 |
| 5 | Kicevo | 47700 | 13628,57 | 2044,29 | 11584,29 | 170,36 | 241,34 | 411,70 | 0,536561 | 220,900 |
| 6 | Krushevo | 9684 | 2766,86 | 415,03 | 2351,83 | 34,59 | 49,00 | 83,58 | 0,518026 | 43,298 |
| 7 | Prilep | 76768 | 21933,71 | 3290,06 | 18643,66 | 274,17 | 388,41 | 662,58 | 0,527954 | 349,812 |
| 8 | Gostivar | 81042 | 23154,86 | 3473,23 | 19681,63 | 289,44 | 410,03 | 699,47 | 0,543557 | 380,201 |
| 9 |  | 86580 | 24737,14 | 3710,57 | 21026,57 | 309,21 | 438,05 | 747,27 | 0,59102 | 441,650 |
| 10 | Skopje | 506926 | 144836,00 | 21725,40 | 123110,60 | 1810,45 | 2564,80 | 4375,25 | 0,58482 | 2558,736 |
| 11 | Kumanovo | 105484 | 30138,29 | 4520,74 | 25617,54 | 376,73 | 533,70 | 910,43 | 0,547523 | 498,480 |
| 12 | Kriva Palanka | 20820 | 5948,57 | 892,29 | 5056,29 | 74,36 | 105,34 | 179,70 | 0,502561 | 90,308 |
| 13 | Sveti Nikole | 18497 | 5284,86 | 792,73 | 4492,13 | 66,06 | 93,59 | 159,65 | 0,51487 | 82,197 |
| 14 | Shtip | 47796 | 13656,00 | 2048,40 | 11607,60 | 170,70 | 241,83 | 412,53 | 0,557689 | 230,061 |
| 15 | Veles | 55108 | 15745,14 | 2361,77 | 13383,37 | 196,81 | 278,82 | 475,63 | 0,527339 | 250,821 |
| 16 | Kocani | 38092 | 10883,43 | 1632,51 | 9250,91 | 136,04 | 192,73 | 328,77 | 0,554843 | 182,416 |
| 17 | Radovish | 28244 | 8069,71 | 1210,46 | 6859,26 | 100,87 | 142,90 | 243,77 | 0,564386 | 137,582 |
| 18 | Regotino | 19212 | 5489,14 | 823,37 | 4665,77 | 68,61 | 97,20 | 165,82 | 0,537121 | 89,064 |
| 19 | Kavadaci | 38741 | 11068,86 | 1660,33 | 9408,53 | 138,36 | 196,01 | 334,37 | 0,529267 | 176,972 |
| 20 | Strumica | 54676 | 15621,71 | 2343,26 | 13278,46 | 195,27 | 276,63 | 471,91 | 0,529578 | 249,911 |
| 21 | Gevgelija | 22988 | 6568,00 | 985,20 | 5582,80 | 82,10 | 116,31 | 198,41 | 0,577857 | 114,652 |
|  | Σ | 1492410 |  |  |  |  |  |  |  |  |

Table 3. Traffic between cities with SEI included.

|  | City | 1 Struga | 2 Ohrid | 3 Bitola | 4 Debar | 5 Kicevo | 6 Krushevo | 7 Prilep | 8 Gostivar | 9 Tetovo | 10 Skopje | 11 Kumanovo | 12 Kriva Palanka | 13 Sveti Nikole | 14 Shtip | 15 Veles | 16 Kocani | 17 Radovish | 18 Negotino | 19 Kavadarci | 20 Strumica | 21 Gevgelija |
|---|---|---|---|---|---|---|---|---|---|---|---|---|---|---|---|---|---|---|---|---|---|---|
| 1 | Struga |  | 11,17 | 18,90 | 3,82 | 9,38 | 1,84 | 14,85 | 16,15 | 18,75 | 108,66 | 21,17 | 3,83 | 3,49 | 9,77 | 10,65 | 7,75 | 5,84 | 3,78 | 7,52 | 10,6 | 4,87 |
| 2 | Ohrid | 11,17 |  | 16,60 | 3,36 | 8,25 | 1,62 | 13,1 | 14,2 | 16,5 | 95,6 | 18,6 | 3,37 | 3,07 | 8,59 | 9,37 | 6,81 | 5,14 | 3,33 | 6,61 | 9,34 | 4,28 |
| 3 | Bitola | 18,90 | 16,62 |  | 5,75 | 14,12 | 2,77 | 22,36 | 24,30 | 28,23 | 163,54 | 31,86 | 5,77 | 5,25 | 14,70 | 16,03 | 11,66 | 8,79 | 5,69 | 11,31 | 15,97 | 7,33 |
| 4 | Debar | 3,82 | 3,36 | 5,75 |  | 2,89 | 0,57 | 4,58 | 4,98 | 5,78 | 33,5 | 6,53 | 1,18 | 1,08 | 3,01 | 3,28 | 2,39 | 1,8 | 1,17 | 2,32 | 3,27 | 1,50 |
| 5 | Kicevo | 9,38 | 8,25 | 14,12 | 2,89 |  | 1,38 | 11,18 | 12,15 | 14,12 | 81,78 | 15,93 | 2,89 | 2,63 | 7,35 | 8,02 | 5,83 | 4,4 | 2,85 | 5,66 | 7,99 | 3,66 |
| 6 | Krushevo | 1,84 | 1,62 | 2,77 | 0,57 | 1,38 |  | 2,27 | 2,47 | 2,87 | 16,6 | 3,23 | 0,59 | 0,53 | 1,49 | 1,63 | 1,18 | 0,89 | 0,58 | 1,15 | 1,62 | 0,74 |
| 7 | Prilep | 14,85 | 13,07 | 22,36 | 4,58 | 11,18 | 2,27 |  | 19,6 | 22,7 | 131,6 | 25,6 | 4,6 | 4,2 | 11,8 | 12,9 | 9,4 | 7,1 | 4,6 | 9,1 | 12,9 | 5,9 |
| 8 | Gostivar | 16,15 | 14,20 | 24,30 | 4,98 | 12,15 | 2,47 | 19,6 |  | 24 | 138,9 | 27,1 | 4,9 | 4,5 | 12,5 | 13,6 | 9,9 | 7,5 | 4,8 | 9,6 | 13,6 | 6,2 |
| 9 |  | 18,75 | 16,50 | 28,23 | 5,78 | 14,12 | 2,87 | 22,7 | 24,0 |  | 148,4 | 28,92 | 5,239 | 4,769 | 13,35 | 14,55 | 10,58 | 7,98 | 5,16 | 10,27 | 14,50 | 6,65 |
| 10 | Skopje | 108,66 | 95,58 | 163,54 | 33,50 | 81,78 | 16,60 | 131,6 | 138,9 | 148,44 |  | 169,3 | 30,7 | 27,9 | 78 | 85,2 | 62 | 46,7 | 30,3 | 60,1 | 17,7 | 38,9 |
| 11 | Kumanovo | 21,17 | 18,62 | 31,86 | 6,53 | 15,93 | 3,23 | 25,6 | 27,1 | 28,92 | 169,32 |  | 6,38 | 5,81 | 16,26 | 17,73 | 12,89 | 9,72 | 6,3 | 12,51 | 17,66 | 8,1 |
| 12 | Kriva Palanka | 3,83 | 3,37 | 5,77 | 1,18 | 2,89 | 0,59 | 4,6 | 4,9 | 5,24 | 30,67 | 6,38 |  | 1,15 | 3,21 | 3,5 | 2,54 | 1,92 | 1,24 | 2,47 | 3,49 | 1,6 |
| 13 | Sveti Nikole | 3,49 | 3,07 | 5,25 | 1,08 | 2,63 | 0,53 | 4,2 | 4,5 | 4,77 | 27,92 | 5,81 | 1,15 |  | 2,85 | 3,11 | 2,26 | 1,71 | 1,10 | 2,19 | 3,10 | 1,42 |
| 14 | Shtip | 9,77 | 8,59 | 14,70 | 3,01 | 7,35 | 1,49 | 11,8 | 12,5 | 13,35 | 78,14 | 16,26 | 3,21 | 2,851 |  | 8,03 | 5,84 | 4,40 | 2,85 | 5,66 | 8,00 | 3,67 |
| 15 | Veles | 10,65 | 9,37 | 16,03 | 3,28 | 8,02 | 1,63 | 12,9 | 13,6 | 14,55 | 85,20 | 17,73 | 3,50 | 3,109 | 8,03 |  | 6,74 | 5,08 | 3,29 | 6,53 | 9,23 | 4,23 |
| 16 | Kocani | 7,75 | 6,81 | 11,66 | 2,39 | 5,83 | 1,18 | 9,4 | 9,9 | 10,58 | 61,96 | 12,89 | 2,54 | 2,261 | 5,84 | 6,74 |  | 3,51 | 2,27 | 4,51 | 6,37 | 2,92 |
| 17 | Radovish | 5,84 | 5,14 | 8,79 | 1,80 | 4,40 | 0,89 | 7,1 | 7,5 | 7,98 | 46,73 | 9,72 | 1,92 | 1,705 | 4,41 | 5,08 | 3,51 |  | 1,69 | 3,35 | 4,73 | 2,17 |
| 18 | Regotino | 3,78 | 3,33 | 5,69 | 1,17 | 2,85 | 0,58 | 4,6 | 4,8 | 5,17 | 30,25 | 6,30 | 1,24 | 1,10 | 2,85 | 3,29 | 2,27 | 1,69 |  | 2,28 | 3,22 | 1,48 |
| 19 | Kavadaci | 7,52 | 6,61 | 11,31 | 2,32 | 5,66 | 1,15 | 9,1 | 9,6 | 10,27 | 60,11 | 12,51 | 2,47 | 2,193 | 5,67 | 6,53 | 4,52 | 3,35 | 2,28 |  | 6,49 | 2,98 |
| 20 | Strumica | 10,61 | 9,34 | 15,97 | 3,27 | 7,99 | 1,62 | 12,9 | 13,6 | 14,50 | 17,66 | 17,66 | 3,49 | 3,10 | 8,00 | 9,23 | 6,38 | 4,73 | 3,22 | 6,49 |  | 4,20 |
| 21 | Gevgelija | 4,87 | 4,28 | 7,33 | 1,50 | 3,66 | 0,74 | 5,9 | 6,2 | 6,65 | 38,94 | 8,10 | 1,60 | 1,421 | 3,67 | 4,23 | 2,93 | 2,17 | 1,48 | 2,98 | 4,20 |  |

International Journal of Computer Networks & Communications (IJCNC) Vol.6, No.3, May 2014

## 3.2. Algorithms for Finding the Shortest Paths

Dijkstra's algorithm solves the problem of finding the shortest path from a source (point of the graph) to the other nodes (points of the graph) when all weights (scalars) of the edges in the graph are positive (Figure 1) [13,14]. The routing table (Forwarding Database) for this graph is seen below (Table 4) and each column of the table holds data from the neighboring routers (Adjacency Database) [15].

Table 4. The data gathered from the neighboring routers.

| ST | OH | BT | DE | KI | KS | PP | GV | TE | SK | KU | KP | SN | SHT | VE | KO | RA | NG | KV | SU | GE |
|---|---|---|---|---|---|---|---|---|---|---|---|---|---|---|---|---|---|---|---|---|
| OH/13 | ST/13 | KS/38 | ST/40 | DE/35 | PP/25 | KI/53 | KI/32 | GV/25 | KU/25 | SK/25 | KU/52 | KU/35 | SN/25 | SK/42 | SHT/27 | SHT/25 | KV/9 | NG/9 | RA/28 | KV/53 |
| Ki/45 | BT/46 | PP/40 | KI/35 | KS/31 | BT/38 | KS/25 | DE/43 | SK/30 | VE/42 | KP/52 | | VE/22 | VE/33 | SN/22 | | NG/35 | RA/35 | PP/38 | GE/35 | SU/35 |
| DE/40 | | OH/46 | GV/43 | GV/32 | KI/31 | BT/40 | TE/25 | | TE/30 | SN/35 | | SHT/25 | KO/27 | SHT/33 | | SU/28 | VE/37 | GE/53 | | |
| | | | | ST/45 | | KV/38 | | | | | | | RA/25 | NG/37 | | | | | | |
| | | | | PP/53 | | | | | | | | | | | | | | | | |

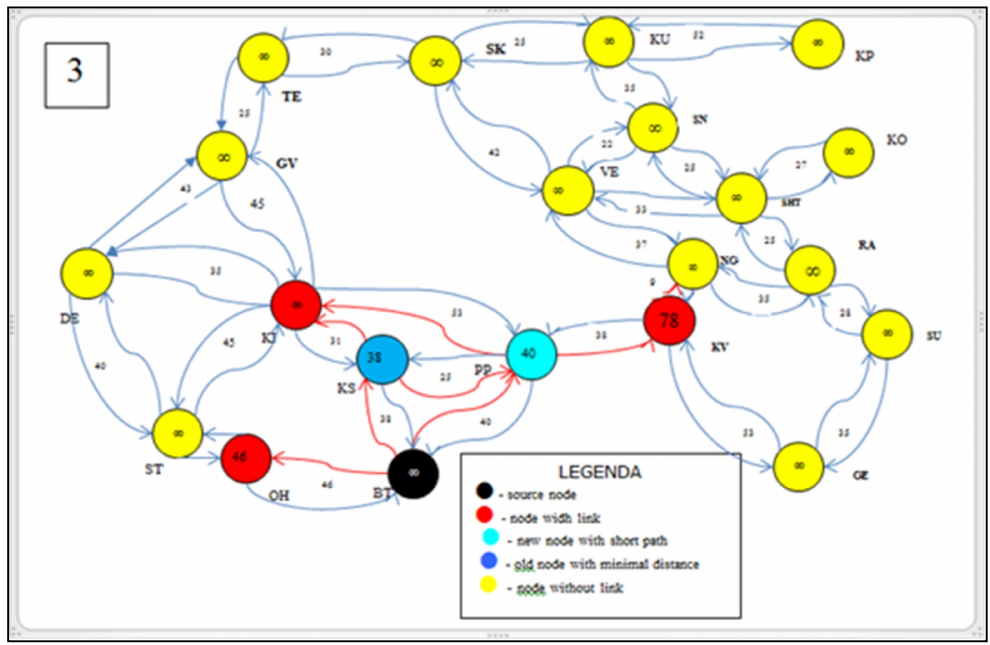

Figure1. Applying the Dijkstra's algorithm in this particular graph
77

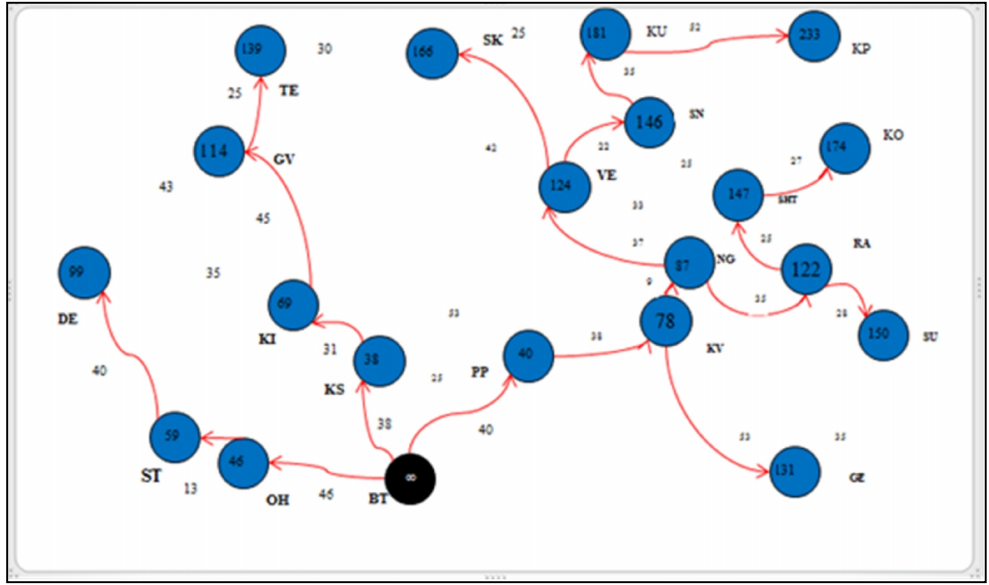

Figure 2. Shortest paths from the BT node (source) with Dijkstra's algorithm

Table 5. Determining shortest paths from the BT node with Dijkstra's algorithm.

| Steps | BT | KS | PP | OH | ST | KI | KV | NG | DE | GV | RA | VE | GE | TE | SN | SHT | SU | SK | KO | KU | KP |
|---|---|---|---|---|---|---|---|---|---|---|---|---|---|---|---|---|---|---|---|---|---|
| 0 | 0/NIL | ∞/NIL | ∞/NIL | ∞/NIL | ∞/NIL | ∞/NIL | ∞/NIL | ∞/NIL | ∞/NIL | ∞/NIL | ∞/NIL | ∞/NIL | ∞/NIL | ∞/NIL | ∞/NIL | ∞/NIL | ∞/NIL | ∞/NIL | ∞/NIL | ∞/NIL | ∞/NIL |
| 1 | 0/NIL | 38/BT | 40/BT | 46/BT | ∞/NIL | ∞/NIL | ∞/NIL | ∞/NIL | ∞/NIL | ∞/NIL | ∞/NIL | ∞/NIL | ∞/NIL | ∞/NIL | ∞/NIL | ∞/NIL | ∞/NIL | ∞/NIL | ∞/NIL | ∞/NIL | ∞/NIL |
| 2 | 0/NIL | 38/BT | 40/BT | 46/BT | ∞/NIL | 69/BT | ∞/NIL | ∞/NIL | ∞/NIL | ∞/NIL | ∞/NIL | ∞/NIL | ∞/NIL | ∞/NIL | ∞/NIL | ∞/NIL | ∞/NIL | ∞/NIL | ∞/NIL | ∞/NIL | ∞/NIL |
| 3 | 0/NIL | 38/BT | 40/BT | 46/BT | ∞/NIL | 69/BT | 78/BT | ∞/NIL | ∞/NIL | ∞/NIL | ∞/NIL | ∞/NIL | ∞/NIL | ∞/NIL | ∞/NIL | ∞/NIL | ∞/NIL | ∞/NIL | ∞/NIL | ∞/NIL | ∞/NIL |
| 4 | 0/NIL | 38/BT | 40/BT | 46/BT | 59/B | 69/BT | 78/BT | ∞/NIL | ∞/NIL | ∞/NIL | ∞/NIL | ∞/NIL | ∞/NIL | ∞/NIL | ∞/NIL | ∞/NIL | ∞/NIL | ∞/NIL | ∞/NIL | ∞/NIL | ∞/NIL |
| 5 | 0/NIL | 38/BT | 40/BT | 46/BT | 59/B | 69/BT | 78/BT | ∞/NIL | 99/BT | ∞/NIL | ∞/NIL | ∞/NIL | ∞/NIL | ∞/NIL | ∞/NIL | ∞/NIL | ∞/NIL | ∞/NIL | ∞/NIL | ∞/NIL | ∞/NIL |
| 6 | 0/NIL | 38/BT | 40/BT | 46/BT | 59/B | 69/BT | 78/BT | ∞/NIL | 99/BT | 114/BT | ∞/NIL | ∞/NIL | ∞/NIL | ∞/NIL | ∞/NIL | ∞/NIL | ∞/NIL | ∞/NIL | ∞/NIL | ∞/NIL | ∞/NIL |
| 7 | 0/NIL | 38/BT | 40/BT | 46/BT | 59/B | 69/BT | 78/BT | 87/BT | 99/BT | 114/BT | ∞/NIL | ∞/NIL | 131/BT | ∞/NIL | ∞/NIL | ∞/NIL | ∞/NIL | ∞/NIL | ∞/NIL | ∞/NIL | ∞/NIL |
| 8 | 0/NIL | 38/BT | 40/BT | 46/BT | 59/B | 69/BT | 78/BT | 87/BT | 99/BT | 114/BT | 122/BT | 124/BT | 131/BT | ∞/NIL | ∞/NIL | ∞/NIL | ∞/NIL | ∞/NIL | ∞/NIL | ∞/NIL | ∞/NIL |
| 9 | 0/NIL | 38/BT | 40/BT | 46/BT | 59/B | 69/BT | 78/BT | 87/BT | 99/BT | 114/BT | 122/BT | 124/BT | 131/BT | ∞/NIL | ∞/NIL | ∞/NIL | ∞/NIL | ∞/NIL | ∞/NIL | ∞/NIL | ∞/NIL |
| 10 | 0/NIL | 38/BT | 40/BT | 46/BT | 59/B | 69/BT | 78/BT | 87/BT | 99/BT | 114/BT | 122/BT | 124/BT | 131/BT | 139/B | ∞/NIL | ∞/NIL | ∞/NIL | ∞/NIL | ∞/NIL | ∞/NIL | ∞/NIL |
| 11 | 0/NIL | 38/BT | 40/BT | 46/BT | 59/B | 69/BT | 78/BT | 87/BT | 99/BT | 114/BT | 122/BT | 124/BT | 131/BT | 139/B | 146/B | 147/BT | 150/BT | ∞/NIL | ∞/NIL | ∞/NIL | ∞/NIL |
| 12 | 0/NIL | 38/BT | 40/BT | 46/BT | 59/B | 69/BT | 78/BT | 87/BT | 99/BT | 114/BT | 122/BT | 124/BT | 131/BT | 139/B | 146/B | 157/BT | 150/BT | 166/B | ∞/NIL | ∞/NIL | ∞/NIL |
| 13 | 0/NIL | 38/BT | 40/BT | 46/BT | 59/B | 69/BT | 78/BT | 87/BT | 99/BT | 114/BT | 122/BT | 124/BT | 131/BT | 139/B | 146/B | 157/BT | 166/B | 166/B | ∞/NIL | ∞/NIL | ∞/NIL |
| 14 | 0/NIL | 38/BT | 40/BT | 46/BT | 59/B | 69/BT | 78/BT | 87/BT | 99/BT | 114/BT | 122/BT | 124/BT | 131/BT | 139/B | 146/B | 157/BT | 166/B | 166/B | ∞/NIL | ∞/NIL | ∞/NIL |
| 15 | 0/NIL | 38/BT | 40/BT | 46/BT | 59/B | 69/BT | 78/BT | 87/BT | 99/BT | 114/BT | 122/BT | 124/BT | 131/BT | 139/B | 146/B | 171/BT | 166/BT | 166/B | ∞/NIL | 181/BT | ∞/NIL |
| 16 | 0/NIL | 38/BT | 40/BT | 46/BT | 59/B | 69/BT | 78/BT | 87/BT | 99/BT | 114/BT | 122/BT | 124/BT | 131/BT | 139/B | 146/B | 147/BT | 166/BT | 166/B | 174/BT | 181/BT | ∞/NIL |
| 17 | 0/NIL | 38/BT | 40/BT | 46/BT | 59/B | 69/BT | 78/BT | 87/BT | 99/BT | 114/BT | 122/BT | 124/BT | 131/BT | 139/B | 146/B | 147/BT | 150/BT | 166/B | 174/BT | 181/BT | ∞/NIL |
| 18 | 0/NIL | 38/BT | 40/BT | 46/BT | 59/B | 69/BT | 78/BT | 87/BT | 99/BT | 114/BT | 122/BT | 124/BT | 131/BT | 139/B | 146/B | 147/BT | 150/BT | 166/B | 174/BT | 182/BT | ∞/NIL |
| 19 | 0/NIL | 38/BT | 40/BT | 46/BT | 59/B | 69/BT | 78/BT | 87/BT | 99/BT | 114/BT | 122/BT | 124/BT | 131/BT | 139/B | 146/B | 147/BT | 150/BT | 166/B | 174/BT | 182/BT | ∞/NIL |
| 20 | 0/NIL | 38/BT | 40/BT | 46/BT | 59/B | 69/BT | 78/BT | 87/BT | 99/BT | 114/BT | 122/BT | 124/BT | 131/BT | 139/B | 146/B | 147/BT | 150/BT | 166/B | 174/BT | 181/BT | 233/BT |
| 21 | 0/NIL | 38/BT | 40/BT | 46/BT | 59/B | 69/BT | 78/BT | 87/BT | 99/BT | 114/BT | 122/BT | 124/BT | 131/BT | 139/B | 146/B | 147/BT | 150/BT | 166/B | 174/BT | 181/BT | 233/BT |

Based on Figure 2 and the final results after applying the Dijkstra's algorithm for finding the shortest path from the BT node to all other cities (Table 5), one can create a table by describing several routes (Table 6).

The Floyd-Warshall algorithm is designed to find the shortest path for all pairs of nodes (points) on a graph (Figure 3). $A_k$ is a matrix of the type $n \times n$, where $A_k[i, j]$ is the weight of the shortest path from $i$ to $j$, which passes through nodes $<= k$ [13,14,16]. In this case we define:

$$A_0[i,j] = \begin{cases} 0 \ldots \text{if } i = j \\ \text{weight of branches from i to j} \ldots \text{for } i \neq j \text{ and} (i,j) \in E \\ \infty \ldots \text{if } i \neq j \text{ and } (i,j) \notin E \end{cases} \quad (6)$$

Now we examine the shortest path $p$ from $i$ to $j$, which passes through nodes $1...k$ – one of two possibilities apply [16]:

- path $p$ does not pass through $k$ and, in this case, the weight of the path is $A_{k-1}[i,j]$;
- path $p$ passes through $k$ and, in this case, the weight of the path is $A_{k-1}[i,k] + A_{k-1}[k,j]$.

Then we have: $A_k[i,j]$ = **min** $(A_{k-1}[i,j], A_{k-1}[i,k] + A_{k-1}[k,j])$.

Based on this particular graph, we create a table in the form of a matrix (Adjacency Matrix). If $i = j$, the $A[i,j]$ element will have value 0 and will represent the diagonal of the matrix. In case there is still no link between nodes the value of the element in the matrix is described by the $\infty$ symbol (Table 7).

Table 6. Shortest paths with Dijkstra's algorithm and corresponding distances.

| Source node | Destination | Distance | Route |
|---|---|---|---|
| Bitola | Krushevo | 38 | Bitola-Krushevo |
| Bitola | Prilep | 40 | Bitola-Prilep |
| Bitola | Ohrid | 46 | Bitola-Ohrid |
| Bitola | Struga | 59 | Bitola-Ohrid-Struga |
| Bitola | Kicevo | 69 | Bitola-Krushevo-Kicevo |
| Bitola | Kavadarci | 78 | Bitola-Prilep-Kavadarci |
| Bitola | Negotino | 87 | Bitola-Prilep-Kavadarci-Negotino |
| Bitola | Debar | 99 | Bitola-Ohrid-Struga-Debar |
| Bitola | Gostivar | 102 | Bitola-Krushevo-Kicevo-Gostivar |
| Bitola | Radovish | 122 | Bitola-Prilep-Kavadarci-Negotino-Radovish |
| Bitola | Veles | 124 | Bitola-Prilep-Kavadarci-Negotino-Veles |
| Bitola | Tetovo | 127 | Bitola-Krushevo-Kicevo-Gostivar-Tetovo |
| Bitola | Gevgelija | 131 | Betola-Prilep-Kavadarci-Gevgelija |
| Bitola | Sveti Nikole | 146 | Bitola-Prilep-Kavadarci-Negotino-Veles-Sveti Nikole |
| Bitola | Shtip | 147 | Bitola-Prilep-Kavadarci-Negotino-Radovish-Shtip |
| Bitola | Strumica | 150 | Bitola-Prilep-Kavadarci-Negotino-Radovish-Strumica |
| Bitola | Skopje | 157 | Bitola-Krushevo-Kicevo-Gostivar-Tetovo-Skopje |
| Bitola | Kocani | 174 | Bitola-Prilep-Kavadarci-Negotino-Radovish-Shtip-Kocani |
| Bitola | Kumanovo | 181 | Bitola-Prilep-Kavadarci-Negotino-Veles-Sveti Nikole-Kumanovo |
| Bitola | Kriva Palanka | 233 | Bitola-Prilep-Kavadarci-Negotino-Veles-Sveti Nikole-Kumanovo-Kriva Palanka |

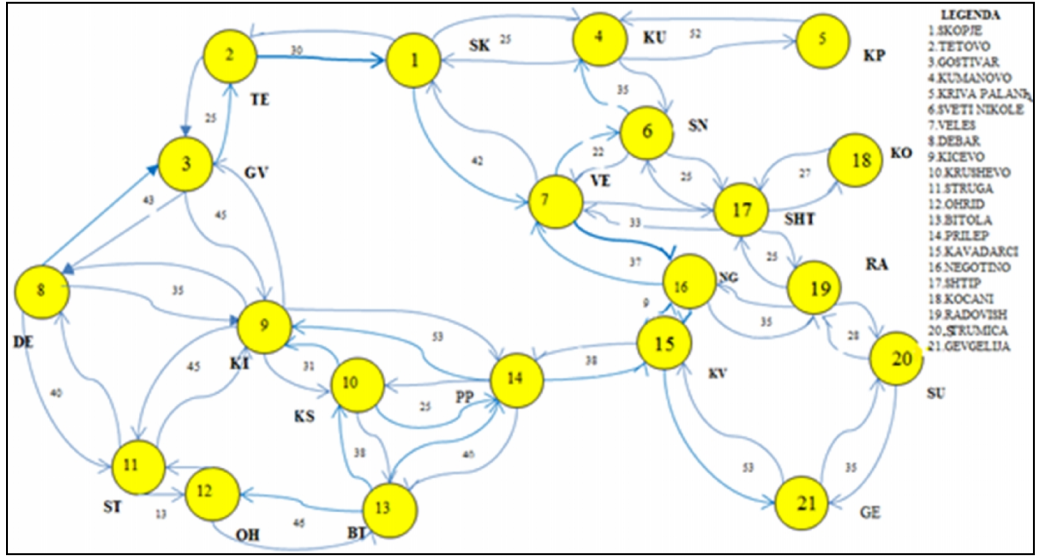

Figure 3. Applying the Floyd-Warshall algorithm in this particular graph

For example, $p$ passes through $k$ and these paths have been found: $p_{42}=30+25=55$, $p_{24}=25+30=55$, $p_{72}=30+42=72$, $p_{27}=72$, $p_{74}=67$, $p_{47}=67$.

Table 7. Determining shortest paths with the Floyd-Warshall algorithm.

| | 1 | 2 | 3 | 4 | 5 | 6 | 7 | 8 | 9 | 10 | 11 | 12 | 13 | 14 | 15 | 16 | 17 | 18 | 19 | 20 | 21 |
|---|---|---|---|---|---|---|---|---|---|---|---|---|---|---|---|---|---|---|---|---|---|
| 1 | 0 | 30 | ∞ | 25 | ∞ | ∞ | 42 | ∞ | ∞ | ∞ | ∞ | ∞ | ∞ | ∞ | ∞ | ∞ | ∞ | ∞ | ∞ | ∞ | ∞ |
| 2 | 30 | 0 | 25 | 55 | ∞ | ∞ | 72 | ∞ | ∞ | ∞ | ∞ | ∞ | ∞ | ∞ | ∞ | ∞ | ∞ | ∞ | ∞ | ∞ | ∞ |
| 3 | ∞ | 25 | 0 | ∞ | ∞ | ∞ | ∞ | 43 | 45 | ∞ | ∞ | ∞ | ∞ | ∞ | ∞ | ∞ | ∞ | ∞ | ∞ | ∞ | ∞ |
| 4 | 25 | 55 | ∞ | 0 | 52 | 35 | 67 | ∞ | ∞ | ∞ | ∞ | ∞ | ∞ | ∞ | ∞ | ∞ | ∞ | ∞ | ∞ | ∞ | ∞ |
| 5 | ∞ | ∞ | ∞ | 52 | 0 | ∞ | ∞ | ∞ | ∞ | ∞ | ∞ | ∞ | ∞ | ∞ | ∞ | ∞ | ∞ | ∞ | ∞ | ∞ | ∞ |
| 6 | ∞ | ∞ | ∞ | 35 | ∞ | 0 | 22 | ∞ | ∞ | ∞ | ∞ | ∞ | ∞ | ∞ | ∞ | ∞ | 25 | ∞ | ∞ | ∞ | ∞ |
| 7 | 42 | 72 | ∞ | 67 | ∞ | 22 | 0 | ∞ | ∞ | ∞ | ∞ | ∞ | ∞ | ∞ | ∞ | 37 | 33 | ∞ | ∞ | ∞ | ∞ |
| 8 | ∞ | ∞ | 43 | ∞ | ∞ | ∞ | ∞ | 0 | 35 | ∞ | 40 | ∞ | ∞ | ∞ | ∞ | ∞ | ∞ | ∞ | ∞ | ∞ | ∞ |
| 9 | ∞ | ∞ | 45 | ∞ | ∞ | ∞ | ∞ | 35 | 0 | 31 | 45 | ∞ | | 53 | ∞ | ∞ | ∞ | ∞ | ∞ | ∞ | ∞ |
| 10 | ∞ | ∞ | ∞ | ∞ | ∞ | ∞ | ∞ | ∞ | 31 | 0 | ∞ | 38 | 25 | ∞ | ∞ | ∞ | ∞ | ∞ | ∞ | ∞ | ∞ |
| 11 | ∞ | ∞ | ∞ | ∞ | ∞ | ∞ | ∞ | 40 | 45 | ∞ | 0 | 13 | | ∞ | ∞ | ∞ | ∞ | ∞ | ∞ | ∞ | ∞ |
| 12 | ∞ | ∞ | ∞ | ∞ | ∞ | ∞ | ∞ | ∞ | ∞ | ∞ | 13 | 0 | 46 | | ∞ | ∞ | ∞ | ∞ | ∞ | ∞ | ∞ |
| 13 | ∞ | ∞ | ∞ | ∞ | ∞ | ∞ | ∞ | ∞ | ∞ | 38 | ∞ | 46 | 0 | 40 | ∞ | ∞ | ∞ | ∞ | ∞ | ∞ | ∞ |
| 14 | ∞ | ∞ | ∞ | ∞ | ∞ | ∞ | ∞ | ∞ | 53 | 25 | ∞ | ∞ | 40 | 0 | 38 | ∞ | ∞ | ∞ | ∞ | ∞ | ∞ |
| 15 | ∞ | ∞ | ∞ | ∞ | ∞ | ∞ | ∞ | ∞ | ∞ | ∞ | ∞ | ∞ | ∞ | 38 | 0 | 9 | ∞ | ∞ | ∞ | ∞ | 53 |
| 16 | ∞ | ∞ | ∞ | ∞ | ∞ | ∞ | 37 | ∞ | ∞ | ∞ | ∞ | ∞ | ∞ | 9 | 0 | ∞ | ∞ | 35 | ∞ | ∞ | ∞ |
| 17 | ∞ | ∞ | ∞ | ∞ | ∞ | 25 | 33 | ∞ | ∞ | ∞ | ∞ | ∞ | ∞ | ∞ | ∞ | 0 | 27 | 25 | ∞ | ∞ | ∞ |
| 18 | ∞ | ∞ | ∞ | ∞ | ∞ | ∞ | ∞ | ∞ | ∞ | ∞ | ∞ | ∞ | ∞ | ∞ | ∞ | 27 | 0 | | ∞ | ∞ | ∞ |
| 19 | ∞ | ∞ | ∞ | ∞ | ∞ | ∞ | ∞ | ∞ | ∞ | ∞ | ∞ | ∞ | ∞ | ∞ | 35 | 25 | ∞ | 0 | 28 | ∞ | ∞ |
| 20 | ∞ | ∞ | ∞ | ∞ | ∞ | ∞ | ∞ | ∞ | ∞ | ∞ | ∞ | ∞ | ∞ | ∞ | ∞ | ∞ | ∞ | 28 | 0 | 35 | ∞ |
| 21 | ∞ | ∞ | ∞ | ∞ | ∞ | ∞ | ∞ | ∞ | ∞ | ∞ | ∞ | ∞ | ∞ | 53 | ∞ | ∞ | ∞ | ∞ | 35 | 0 | |

International Journal of Computer Networks & Communications (IJCNC) Vol.6, No.3, May 2014

Table 8. Shortest paths between all pairs in the graph with the Floyd-Warshall algorithm.

|    | 1 | 2 | 3 | 4 | 5 | 6 | 7 | 8 | 9 | 10 | 11 | 12 | 13 | 14 | 15 | 16 | 17 | 18 | 19 | 20 | 21 |
|---|---|---|---|---|---|---|---|---|---|---|---|---|---|---|---|---|---|---|---|---|---|
| 1 | 0 | 30 | 55 | 25 | 77 | 60 | 42 | 98 | 100 | 131 | 138 | 151 | 166 | 126 | 88 | 79 | 75 | 102 | 100 | 128 | 141 |
| 2 | 30 | 0 | 25 | 55 | 107 | 90 | 72 | 68 | 70 | 101 | 108 | 121 | 139 | 123 | 118 | 109 | 105 | 132 | 130 | 158 | 171 |
| 3 | 55 | 25 | 0 | 80 | 132 | 115 | 97 | 43 | 45 | 76 | 83 | 96 | 114 | 98 | 136 | 134 | 130 | 167 | 155 | 183 | 189 |
| 4 | 25 | 55 | 80 | 0 | 52 | 35 | 57 | 123 | 125 | 156 | 163 | 176 | 181 | 141 | 103 | 94 | 60 | 87 | 85 | 113 | 148 |
| 5 | 77 | 107 | 132 | 52 | 0 | 87 | 109 | 175 | 177 | 208 | 215 | 228 | 233 | 193 | 155 | 146 | 112 | 139 | 137 | 165 | 208 |
| 6 | 60 | 90 | 115 | 35 | 87 | 0 | 22 | 158 | 159 | 131 | 198 | 192 | 146 | 106 | 68 | 59 | 25 | 52 | 50 | 78 | 113 |
| 7 | 42 | 72 | 97 | 57 | 109 | 22 | 0 | 140 | 137 | 109 | 180 | 170 | 124 | 84 | 46 | 37 | 33 | 60 | 58 | 86 | 99 |
| 8 | 98 | 68 | 43 | 123 | 175 | 158 | 140 | 0 | 35 | 66 | 40 | 53 | 99 | 88 | 126 | 135 | 173 | 200 | 170 | 198 | 179 |
| 9 | 100 | 70 | 45 | 125 | 177 | 159 | 137 | 35 | 0 | 31 | 45 | 58 | 69 | 53 | 91 | 100 | 160 | 187 | 135 | 163 | 144 |
| 10 | 131 | 101 | 76 | 156 | 208 | 131 | 109 | 66 | 31 | 0 | 76 | 89 | 38 | 25 | 63 | 72 | 132 | 159 | 107 | 135 | 116 |
| 11 | 138 | 108 | 83 | 163 | 215 | 198 | 180 | 40 | 45 | 76 | 0 | 13 | 59 | 98 | 136 | 145 | 205 | 232 | 180 | 208 | 189 |
| 12 | 151 | 121 | 96 | 176 | 228 | 192 | 170 | 53 | 58 | 84 | 13 | 0 | 46 | 86 | 124 | 133 | 193 | 220 | 168 | 196 | 177 |
| 13 | 166 | 139 | 114 | 181 | 233 | 146 | 124 | 99 | 69 | 38 | 59 | 46 | 0 | 40 | 78 | 87 | 147 | 174 | 122 | 150 | 131 |
| 14 | 126 | 123 | 98 | 141 | 193 | 106 | 84 | 88 | 53 | 25 | 98 | 86 | 40 | 0 | 38 | 47 | 107 | 134 | 82 | 110 | 91 |
| 15 | 88 | 118 | 136 | 103 | 155 | 68 | 46 | 126 | 91 | 63 | 136 | 124 | 78 | 38 | 0 | 9 | 69 | 96 | 44 | 72 | 53 |
| 16 | 79 | 109 | 134 | 94 | 146 | 59 | 37 | 135 | 100 | 72 | 145 | 133 | 87 | 47 | 9 | 0 | 60 | 87 | 35 | 63 | 62 |
| 17 | 75 | 105 | 130 | 60 | 112 | 25 | 33 | 173 | 160 | 132 | 205 | 193 | 147 | 107 | 69 | 60 | 0 | 27 | 25 | 53 | 88 |
| 18 | 102 | 132 | 157 | 87 | 139 | 52 | 60 | 200 | 187 | 159 | 232 | 220 | 174 | 134 | 96 | 87 | 27 | 0 | 52 | 80 | 115 |
| 19 | 100 | 130 | 155 | 85 | 137 | 50 | 58 | 170 | 135 | 107 | 180 | 168 | 122 | 82 | 44 | 35 | 25 | 52 | 0 | 28 | 63 |
| 20 | 128 | 158 | 183 | 113 | 165 | 78 | 86 | 198 | 163 | 135 | 208 | 196 | 150 | 110 | 72 | 63 | 53 | 80 | 28 | 0 | 35 |
| 21 | 141 | 171 | 189 | 148 | 200 | 113 | 99 | 179 | 144 | 116 | 189 | 177 | 131 | 91 | 53 | 62 | 88 | 115 | 63 | 35 | 0 |

Based on the results given in Table 8, obtained by applying the Floyd-Warshall algorithm, and the map of cities with numbers ranging from 1 to 21 which correspond to symbols (SK, TE, GO, KU, KP, SN, VE, DE, KI, KS, ST, OH, BT, PP, KV, NG, SHT, KO, RA, SU, GE), one can check whether the shortest path between those cities was found and can mark the route (Fig. 4).

Example:  8(DE) → 18(KO) with shortest path 200;
13(BT) → 5(KP) with shortest path 233;
2(TE) → 21(GE) with shortest path 171.

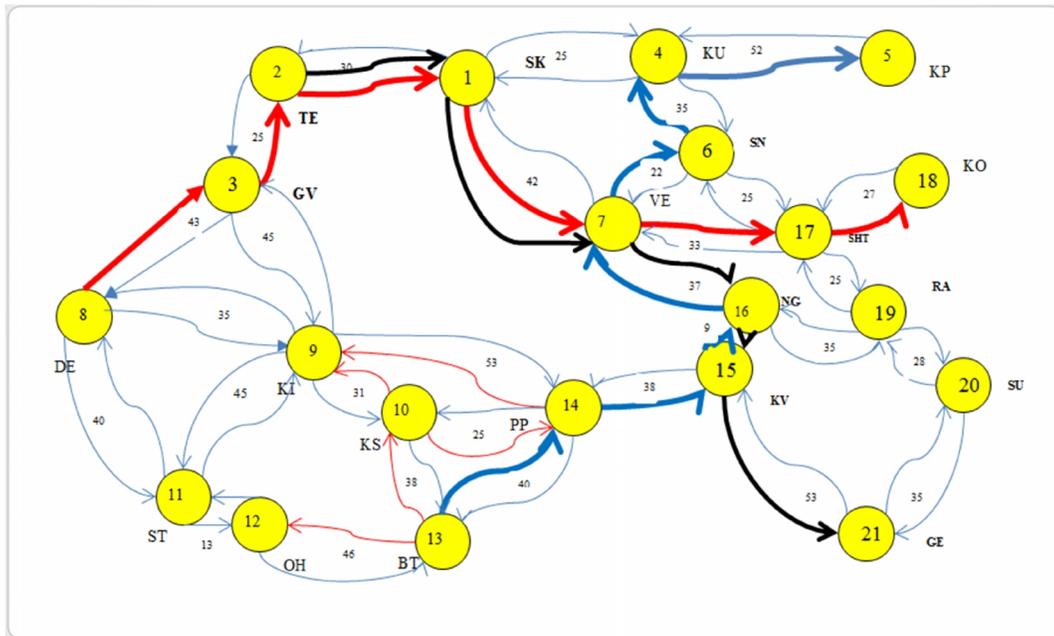

Figure 4. Marking the shortest path for three node pairs

International Journal of Computer Networks & Communications (IJCNC) Vol.6, No.3, May 2014

Since the graph has 21 nodes, there are some 200 routes which describe the shortest path in one direction between all pairs of nodes on this graph. From Table 8, all the values of the shortest paths in both directions can be read (e.g. Gevgelija – Strumica = 35, Strumica – Gevgelija = 35). Below we describe several routes (Table 9).

Table 9. Shortest paths with the Floyd-Warshall Algorithm.

| Node1 | Node2 | Distance | Route with shortest path (Floyd- Warshal) |
|---|---|---|---|
| Gevgelija(21) | Strumica(20) | 35 | Gevgelija-Strumica |
| Gevgelija(21) | Radovish(19) | 63 | Gevgelija-Strumica-Radovish |
| Gevgelija(21) | Shtip(17) | 88 | Gevgelija-Strumica-Radovish-Shtip |
| Gevgelija(21) | Kocani(18) | 115 | Gevgelija-Strumica-Radovish-Shtip-Kocani |
| Gevgelija(21) | Sveti Nikole(6) | 113 | Gevgelija-Strumica-Radovish-Shtip-Sveti Nikole |
| Gevgelija(21) | Kumanovo(4) | 148 | Gevgelija-Strumica-Radovish-Shtip-Sveti Nikole –Kumanovo |
| Gevgelija(21) | Kriva Palanka(5) | 200 | Gevgelija-Strumica-Radovish-Shtip-Sveti Nikole –Kumanovo-Kriva Palanka |
| Gevgelija(21) | Negotino(16) | 62 | Gevgelija-Kavadarci-Negotino |
| Gevgelija(21) | Kavadarci(15) | 53 | Gevgelija-Kavadarci |
| Gevgelija(21) | Veles(7) | 99 | Gevgelija-Kavadarci-Negotino-Veles |
| Gevgelija(21) | Skopje(1) | 141 | Gevgelija-Kavadarci-Negotino-Veles-Skopje |

### 3.3. Calculating Traffic between the Nodes of the Network

To create a bandwidth of each link of the network one needs to calculate traffic (in Erlangs) for each link. This is done by passing through several stages:

- Previous calculation of the traffic of cities including SEI should be done.
- Then all the traffic between the cities is calculated and a traffic matrix between all pairs of nodes is created.
- Algorithms for finding the shortest path (e.g. Floyd-Warshall) between all pairs of nodes (cities) are applied.

Based on those routes, carried traffic in Erlangs is calculated. There are cases when a link is more loaded because there is a possibility for several routes to pass through that particular link, which determine shortest paths of a few pairs of nodes. As a result, several different traffic values for different pairs of cities are summed, and such specific links have increased traffic in Erlangs. Similarly, there are parts of routes which are not used in the determination of the traffic, because routes that do not pass across the link where traffic is calculated will simply have "zero" values. An example of the calculation of traffic between Tetovo and Skopje nodes is presented in Table 10.

By following the same methodology, the traffic for 30 pairs of nodes is calculated (Table 11): the maximum traffic (in Erlangs) is present at the Skopje-Tetovo link (908.74), while the minimum value is in the Prilep-Krushevo link (9.60).

82

Table 10. Traffic (in Erlangs): Tetovo-Skopje

| Link | Shortest Path (km) | Run / not across Tetovo-Skopje link | Comunication (with SEI included) |
|---|---|---|---|
| Tetovo-Skopje | 30 | YES | 148.4 |
| Tetovo-Gostivar | 25 | NO | 0 |
| Tetovo-Kumanovo | 55 | YES | 28.92 |
| Tetovo-Kriva Palanka | 107 | YES | 5.23 |
| Tetovo-Sveti Nikole | 90 | YES | 4.76 |
| Tetovo-Veles | 72 | YES | 14.55 |
| Tetovo-Debar | 68 | NO | 0 |
| Tetovo-Kicevo | 70 | NO | 0 |
| Tetovo-Krushevo | 101 | NO | 0 |
| Tetovo-Struga | 108 | NO | 0 |
| Tetovo-Ohrid | 121 | NO | 0 |
| Tetovo-Bitola | 139 | NO | 0 |
| Tetovo-Prilep | 123 | NO | 0 |
| Tetovo-Kavadarci | 118 | YES | 10.27 |
| Tetovo-Negotino | 109 | YES | 5.17 |
| Tetovo-Shtip | 105 | YES | 13.35 |
| Tetovo-Kocani | 132 | YES | 10.58 |
| Tetovo-Radovish | 130 | YES | 7.98 |
| Tetovo-Strumica | 158 | YES | 14.50 |
| Tetovo-Gevgelija | 171 | YES | 6.65 |
| Skopje-Gostivar | 55 | YES | 138.9 |
| Skopje-Kumanovo | 25 | NO | 0 |
| Skopje-Kriva Palanka | 77 | NO | 0 |
| Skopje-Sveti Nikole | 60 | NO | 0 |
| Skopje-Veles | 42 | NO | 0 |
| Skopje- Debar | 98 | YES | 33.50 |
| Skopje-Kicevo | 100 | YES | 81.78 |
| Skopje-Krushevo | 131 | YES | 16.6 |
| Skopje-Struga | 138 | YES | 108.6 |
| Skopje-Ohrid | 151 | YES | 95.5 |
| Skopje-Bitola | 166 | YES | 163.5 |
| Skopje-Prilep | 126 | NO | 0 |
| Skopje-Kavadarci | 88 | NO | 0 |
| Skopje-Negotino | 79 | NO | 0 |
| Skopje-Shtip | 75 | NO | 0 |
| Skopje-Kocani | 102 | NO | 0 |
| Skopje-Radovish | 100 | NO | 0 |
| Skopje-Strumica | 128 | NO | 0 |
| Skopje-Gevgelija | 141 | NO | 0 |
| TETOVO-SKOPJE (Erlangs) | | | **908.74** |

International Journal of Computer Networks & Communications (IJCNC) Vol.6, No.3, May 2014

Table 11. Traffic (in Erlangs) fordifferent pairs of nodes.

|    | **Node1-Node2**          | **Erlangs** |
|----|--------------------------|-------------|
| 1  | Skopje-Tetovo            | **908.74**  |
| 2  | Skopje-Kumanovo          | **319**     |
| 3  | Skopje-Veles             | **756**     |
| 4  | Tetovo-Gostivar          | **383.2**   |
| 5  | Gostivar-Kicevo          | **235.6**   |
| 6  | Gostivar-Debar           | **91.1**    |
| 7  | Kicevo-Debar             | **15.8**    |
| 8  | Kicevo-Struga            | **92.6**    |
| 9  | Kicevo-Krushevo          | **41.6**    |
| 10 | Ohrid-Struga             | **190.9**   |
| 11 | Struga-Debar             | **193.6**   |
| 12 | Bitola-Ohrid             | **89.3**    |
| 13 | Bitola-Krushevo          | **68.7**    |
| 14 | Bitola-Prilep            | **320.6**   |
| 15 | Prilep-Kicevo            | **117.3**   |
| 16 | Prilep-Krushevo          | **9.6**     |
| 17 | Prilep-Kavadarci         | **286.4**   |
| 18 | Kavadarci-Negotino       | **133.9**   |
| 19 | Kavadarci-Gevgelija      | **88**      |
| 20 | Veles-Negotino           | **118.5**   |
| 21 | Shtip-Veles              | **130.7**   |
| 22 | Kocani-Shtip             | **175.2**   |
| 23 | Shtip-Radovish           | **161.2**   |
| 24 | Shtip-Sveti Nikole       | **30**      |
| 25 | Veles-Sveti Nikole       | **42.6**    |
| 26 | SvetiNikole-Kumanovo     | **206.7**   |
| 27 | Kumanovo-Kriva Palanka   | **87.9**    |
| 28 | Radovish-Negotino        | **48.3**    |
| 29 | Radovish-Strumica        | **172.4**   |
| 30 | Strumica-Gevgelija       | **23.9**    |

### 3.4. Calculating Bandwidth between the Nodes of the Network

The term "bandwidth" for communication networks means *amount of data over time that can be exchanged from point A to point B*[bits per second]. This range depends on the amount of the traffic that passes through the link – when the calculated value is multiplied by 64 kbps, one could get the bandwidth of that particular link, as seen in Table 12 below:

$$Bandwidth = Erlangs \cdot 64\ kbps \quad\quad (7)$$

84…84…84848484848484848484848484848484848484

Table12. Estimated link bandwidth.

|   | LINK | TRAFFIC (Erlangs) | BANDWIDTH (bits per sec) | Mbps |
|---|---|---|---|---|
| 1 | Skopje-Tetovo | **908.74** | 58159360 | 58.1593 |
| 2 | Skopje-Veles | **756** | 48384000 | 48.384 |
| 3 | Tetovo-Gostivar | **383.2** | 24524800 | 24.5248 |
| 4 | Bitola-Prilep | **320.6** | 20518400 | 20.5184 |
| 5 | Skopja-Kumanovo | **319** | 20416000 | 20.416 |
| 6 | Prilep-Kavadarci | **286.4** | 18329600 | 18.3296 |
| 7 | Gostivar-Kicevo | **235.6** | 15078400 | 15.0784 |
| 8 | Sveti Nikole-Kumanovo | **206.7** | 13228800 | 13.2288 |
| 9 | Struga-Debar | **193.6** | 12390400 | 12.3904 |
| 10 | Ohrid-Struga | **190.9** | 12217600 | 12.2176 |
| 11 | Kocani-Shtip | **175.2** | 11212800 | 11.2128 |
| 12 | Radovish-Strumica | **172.4** | 11033600 | 11.0336 |
| 13 | Shtip-Radovish | **161.2** | 10316800 | 10.3168 |
| 14 | Kavadarci-Negotino | **133.9** | 8569600 | 8.5696 |
| 15 | Shtip-Veles | **130.7** | 8364800 | 8.3648 |
| 16 | Veles-Negotino | **118.5** | 7584000 | 7.584 |
| 17 | Prilep-Kicevo | **117.3** | 7507200 | 7.5072 |
| 18 | Kicevo-Struga | **92.6** | 5926400 | 5.9264 |
| 19 | Gostivar-Debar | **91.1** | 5830400 | 5.8304 |
| 20 | Bitola-Ohrid | **89.3** | 5715200 | 5.7152 |
| 21 | Kavadarci-Gevgelija | **88** | 5632000 | 5.632 |
| 22 | Kumanovo-Kriva Palanka | **87.9** | 5625600 | 5.6256 |
| 23 | Bitola-Krushevo | **68.7** | 4396800 | 4.3968 |
| 24 | Radovish-Negotino | **48.3** | 3091200 | 3.0912 |
| 25 | Veles-Sveti Nikole | **42.6** | 2726400 | 2.7264 |
| 26 | Kicevo-Krushevo | **41.6** | 2662400 | 2.6624 |
| 27 | Shtip-Sveti Nikole | **30** | 1920000 | 1.92 |
| 28 | Strumica-Gevgelija | **23.9** | 1529600 | 1.5296 |
| 29 | Kicevo-Debar | **15.8** | 1011200 | 1.0112 |
| 30 | Prilep-Krushevo | **9.6** | 614400 | 0.6144 |

Finally, based on the previous calculations, the bandwidth is determined, along with the cable or a medium that can support that amount of traffic. These proportions are seen in the image below (Figure 5).

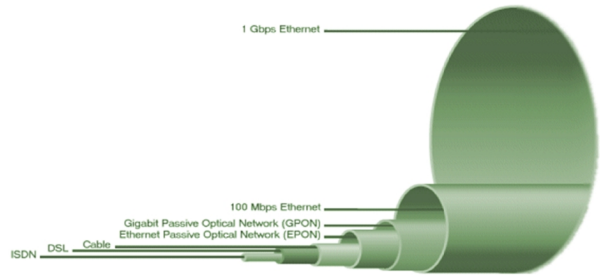

Figure 5. Cable proportions depending on bandwidth[1]

## 4. COST-MODELING

Prices for trunk segments of rented lines are formed on the basis of the reference price model of rented line (LLPC) with the recommendation of the European Commission EC 2005/951/1 final [17]. Rented lines are intended for use by small and large firms, where firms' work depends on the need for a permanent presence on the Internet and high speed data transfers. Advantages of rented lines are: high speed, reliability, a larger number of Internet users at the same time and the ability to be continuously present on the Internet.

Prices of rented lines are quite high because they are paid according to the length and the cost of individual services should also be added, and therefore users are looking for new cheaper rented lines. The retail market users prefer using virtual private networks or xDSL access and Ethernet access. Therefore, xDSL and cable Internet are very interesting for the users in terms of price compared with the prices of rented lines. At the moment of the analysis of the retail market, there were several service providers that use rented lines from AD Makedonski Telekom and offered services. It should be noted that AD Makedonski Telekom has been active in the retail market, providing services for the largest number of users (Table 13) [17].

Table 13. Prices for terminal and trunk segments of leased lines.

| Capacity | Prices in MKD (1 EUR = 61 MKD) |
| --- | --- |
| 64 kbit/s (to 2 km) | 3,733.00 |
| 2 Mbit/s (to 2 km) | 11,383.00 |
| 34 Mbit/s (to 2 km) | 54,590.00 |
| 155 Mbit/s (to 2 km) | 73,807.00 |
| 64 kbit/s (to 5 km) | 4,774.00 |
| 2 Mbit/s (to 5 km) | 15,178.00 |
| 34 Mbit/s (to 5 km) | 58,936.00 |
| 155 Mbit/s (to 5 km) | 81,518.00 |
| 64 kbit/s (to 15 km) | 5,018.00 |
| 2 Mbit/s (to 15 km) | 20,380.00 |
| 34 Mbit/s (to 15 km) | 97,736.00 |
| 155 Mbit/s (to 15 km) | 121,849.00 |
| 64 kbit/s (to 50 km) | 6,059.00 |
| 2 Mbit/s (to 50 km) | 32,987.00 |
| 34 Mbit/s (to 50 km) | 155,387.00 |
| 155 Mbit/s (to 50 km) | 253,613.00 |

---

[1] http://www.technologyuk.net/telecommunications/telecom_principles/bandwidth.shtml (Accessed March, 2014)

International Journal of Computer Networks & Communications (IJCNC) Vol.6, No.3, May 2014

• In the case of ADSL, the price is calculated by a formula where we gather certain variables – the calculation depends on the type of communication network[9]:
- full-duplex;
- half-duplex;
- simplex.

Price for one year of full-duplex is calculated by the formula:
$$\text{cost} = 2 \cdot \text{acc} + 2 \cdot 12 \cdot \text{suba} + 2 \cdot \text{int} + 2 \cdot 12 \cdot \text{subn}$$

Parameters that are part of the overall calculation of price:

- **acc**: (access to network) price of joining the network (paid once), multiplied by two because both directions are paid
- **suba**: (subscription in access to network) price of joining the network (monthly fee), multiplied by 12 months and two directions of communication
- **int**: (initialization) price when the router initializes a link with the server in a radius, multiplied by two because communication is in both directions
- **subn**: (subscription) price of the package by month multiplied by 12 months and two directions of communication

Price for one year of half-duplex is calculated by the formula:

$$\text{cost} = \text{acc} + 2 \cdot 12 \cdot \text{suba} + 2 \cdot \text{int} + 12 \cdot \text{subn}$$

In this case the communication goes in both directions but not simultaneously. **Acc** parameter is not multiplied by two because only one direction has to be paid. It happens to the price of the package **subn** as well. **Int** parameter is multiplied by two because initiation can occur in one of two extremes on the link.

Price for one year of simplex is calculated by the formula:

$$\text{cost} = \text{acc} + 12 \cdot \text{suba} + \text{int} + 12 \cdot \text{subn}$$

In this case the communication takes place only in one direction. We can see this in the formula for calculating the price for one year.

In cases where line is rented, the price for one year for the full-duplex is:
$$\text{cost} = 2 \cdot (\text{int} + 12 \cdot \text{sub})$$

For the half-duplex and simplex the price for one year is calculated by:
$$\text{cost} = \text{int} + 12 \cdot \text{sub}$$

• For the Asynchronous Transfer Mode (ATM) technology, the calculation of the price for full-duplex from point A to point B is calculated by:
$$NA(A) = INS_A + SUB_A \cdot 12$$

The $INS_A$ parameter determines the price of the connection point A, and the $SUB_A$ parameter is the price for a monthly fee.

The following parameters are for calculating the price from point B to the permanent network:
$$NA(B) = INS_B + SUB_B \cdot 12$$

The price of the permanent network that is used is multiplied by 2 because both directions are used. It is calculated by the formula:
$$PVCC = 2 \cdot (INS + SUB \cdot 12)$$

The total cost from point A to point B is the sum price from point A to the permanent network, the price of the permanent network PVCC and the price of the permanent network to point B:
$$TOTAL\_COST = NA(A) + NA(B) + PVCC$$

The price for half-duplex from point A to point B is calculated using the formula:
$$NA(A) = INS_A + SUB_A \cdot 12$$

Similarly,
$$NA(B) = INS_B + SUB_B \cdot 12$$

The price of the permanent network that is used as a proxy of the two points is:
$$PVCC = INS + SUB \cdot 12$$

Again, the total cost from point A to point B is a sum price from point A to the permanent network, the price of the permanent network PVCC and price from the permanent network to point B:
$$TOTAL\_COST = NA(A) + NA(B) + PVCC$$

The calculation for the simplex from point A to point B is done using the formula:
$$NA(A) = INS_A + SUB_A \cdot 12$$
$$PVCC = INS + SUB \cdot 12$$

The total price calculating from point A to point B is a sum price from point A to the permanent network and the price of the permanent network PVCC – since simplex communication link is in one direction one does not have the cost and the price from the permanent network to point B:
$$TOTAL\_COST = NA(A) + PVCC$$

## 5. CONCLUSIONS

The calculation of relevant parameter set provides indicators that help in the process of planning and modeling of WAN networks with adequate capacity and minimum price. They include: the communication matrix for each city, number of households, number of network users, total traffic for any city, the traffic matrix between all cities and shortest paths between the nodes of the graph, which are determined by using appropriate algorithms such as Dijkstra's and/or Floyd-Warshall.

As a tangible contribution, the calculation of a so-called *socioeconomic indicator* (SEI) is made up of a dozen variables which are regularly used in official statistics to illustrate patterns of behavior and outcomes, and to support and develop policies. This again is not enough to estimate the load of an edge of the graph, which connects two adjacent nodes – there might be several shortest paths that pass through an edge. This is how the carried traffic (in Erlangs) for an edge is calculated, as well as the edge bandwidth, assuming that each unit of traffic measurement (Erlang) is equivalent to 64 kbps.

The rationale behind the identification and the evaluation of these parameters is very straightforward: to develop a model which will determine the exact cost of the network that performs within the limits set by the demands of prospective users, by including some socioeconomic variables that capture different levels of technological development and different patterns of behavior.

## REFERENCES

[1]     Marcus, J.S. (1999), *Designing Wide Area Networks and Internetworks*, Addison-Wesley.

[2]     Norris, M. & Pretty, S. (2000), *Designing the Total Area Network: Intranets, VPNs and Enterprise Networks Explained*, Wiley-BT Series.

[3]     Mukherjee, B., Banerjee, D., Ramamurthy, S. & Mukherjee, A. (1996), "Some principles for designing a wide-area WDM optical network", *IEEE/ACM Transactions on Networking*, Vol. 4, No. 5, pp. 684-696

[4]     Del Giudice, P.S. & Amoza, F.R. (2012), "Designing WAN Topologies under Redundancy Constraints", *Optical Fiber Communications and Devices*, Yasin, M. (Ed.), InTech.

[5]     Bigos, W., Cousin, B., Gosselin, S., Le Foll, M. & Nakajima, H. (2007), "Survivable MPLS Over Optical Transport Networks: Cost and Resource Usage Analysis", *IEEE Journal on Selected Areas in Communications,* Vol. 25, No. 5, pp. 949 – 962.

[6]     Chou, W. (2009), "Optimizing the WAN between Branch Offices and the Data Center", *IT Professional*, Vol. 11, No. 4, pp. 24-27.

[7]     Kuribayashi, S. (2013), "Improving Quality of Service and Reducing Power Consumption with WAN accelerator in Cloud Computing Environments", *International Journal of Computer Networks & Communications (IJCNC)*, Vol.5, No.1, pp. 41-52.

[8]     Barney, D. (2011), "Cloud vs. WAN Costs: A Breakdown", *Redmond Magazine*, http://redmondmag.com/articles/2011/12/01/cloud-vs-wan-costs.aspx (Accessed February, 2014)

[9]     Al-Wakeel, S.S. (2009), *"Development of Planning and Cost Models  for Designing A Wide Area Network in Kingdom of Saudi Arabia"*, Research Report #9, Research Center, College of Computer and Information Sciences, King Saud University.

[10]    De Montis, A., Barthelemy, M., Chessa, A. & Vespignani, A. (2007), "The structure of Inter-Urban traffic: A weighted network analysis", *Environment and Planning: B*, Vol. 34, pp. 905-924.

[11]    Taylor, W.J., Zhu, G.X., Dekkers, J. & Marshall, S. (2003), "Socio-Economic Factors Affecting Home Internet Usage Patterns in Central Queensland", *Informing Science Journal*, Vol. 6, Central Queensland University, Rockhampton, Qld, Australia.

[12]    State Statistical Office (2005), *"Total population, households and dwellings according to the territorial organization of the Republic of Macedonia, 2004"*, Skopje, Macedonia.

[13]    Cormen, T.H., Leirserson, C.E., Rivest, R.L. & Stein, C. (2009) *Introduction to Algorithms,* The MIT Press.

[14]    Bollobas, B. (2013), *Modern Graph Theory*, Springer.

[15]    Rhodes, J. (2007), *"Adjacency Matrices in Dijkstra's Shortest Path Algorithm"*, University of North Carolina at Chapel Hill.

[16]    Larsson, C.(2014), *Design of Modern Communication Networks: Methods and Applications*, Academic Press.

[17]    Agency for Electronic Communications (2010),Market Analyses, http://www.aec.mk/index.php ?option=com_content&view=article&id=440&Itemid=81&lang=mk (Accessed March, 2014)

**Authors**

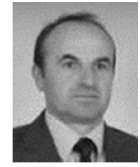

**Basri Ahmedi** received his MSc degree in Computer Science from the South East European University in Tetovo, Republic of Macedonia. He is a teaching and research assistant at the Faculty of Natural Sciences and Mathematics, State University of Tetovo, Republic of Macedonia, and currently works on his PhD thesis at the Department of Computer Science and Engineering at the Faculty of Technical Sciences, St. Clement Ohridski University, Bitola, Republic of Macedonia. His research interests include Computer Networks, Graph Theory and Dynamic Programming.

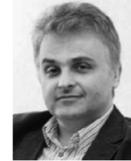

**Pece Mitrevski** received his BSc and MSc degrees in Electrical Engineering and Computer Science, and the PhD degree in Computer Science from the Ss. Cyril and Methodius University in Skopje, Republic of Macedonia. He is currently a full professor and Head of the Department of Computer Science and Engineering at the Faculty of Technical Sciences, St. Clement Ohridski University, Bitola, Republic of Macedonia. His research interests include Computer Networks, Computer Architecture, High Performance Computing, Modeling and Simulation, Performance and Reliability Analysis of Computer Systems and Stochastic Petri Nets. He has published more than 90 papers in journals and refereed conference proceedings and lectured extensively on these topics. He is a member of the IEEE Computer Society and the ACM.